# Radioactive Ion Sources

*T. Stora*
CERN, Geneva, Switzerland

**Abstract**
This chapter provides an overview of the basic requirements for ion sources designed and operated in radioactive ion beam facilities. The facilities where these sources are operated exploit the isotope separation online (ISOL) technique, in which a target is combined with an ion source to maximize the secondary beam intensity and chemical element selectivity. Three main classes of sources are operated, namely surface-type ion sources, arc discharge-type ion sources, and finally radio-frequency-heated plasma-type ion sources.

## 1 Introduction

Radioactive ion sources are used in facilities that deliver radioactive ion beams. These beams are exploited by a large community of physicists active in nuclear physics, astrophysics, atomic physics, material science and life science. The production of the radioactive ions is based on two generic principles: a heavy-ion beam fragmented by a thin target and mass-separated 'in flight' in a series of large-acceptance dipole magnets; and online mass separation of nuclear products diffused out of thick targets impinged by light ion beams, this being referred to as the 'ISOL technique' [1]. In the in-flight technique, the fragments are charged and thus do not require the exploitation of ion sources; while in ISOL production, the fragments diffuse out of a production target or catcher as neutral species.

While ion sources are intimately related to the history and developments of stable particle accelerators, radioactive ion sources are likewise linked to the development of ISOL facilities. In such facilities, ion sources are one of the important links in the full production chain of the secondary radioactive ion beams. The first generation of radioactive ion sources were those developped for stable ions in accelerators such as cyclotrons and mass separators. At a later stage they were adapted to match the specific needs for the production and acceleration of radioisotopes, which are produced with several orders of magnitude less intensities than their stable counterparts, and have, by definition, a finite lifetime before radioactive decay.

The lecture given during the school could not cover all the required aspects in an exhaustive manner. This chapter will introduce some additional aspects such as some historical elements concerned with the development of the sources used in radioactive beam facilities, and some references will be provided to those which need information at an expert level. The chapter is organized in the following manner. In a first part, a brief technical description of a radioactive ion beam facility and the main requirements for a radioactive ion source are introduced. The different classes of ion sources are then reviewed. The sources are split into three main categories; surface-type ion sources, the forced electron beam sources, and finally the RF-heated plasma sources. For each of these categories, historical developments are reported first, followed by the theoretical considerations underlying their operation, and finally the main operational parameters. An outlook for the on-going and future needs in our facilities will finally be reported.

## 2 Radioactive ion beam facilities

The first report of the online production of a radioactive ion beam dates back to the 1950s, when exotic krypton fission fragments (of a few tens of seconds half-life) were produced from large $UO_2$ targets heated and connected to an isotope mass separator [1]. The ion source was of a glow discharge Nielsen type (Fig. 1) [2]. About 15 years after, a number of so-called first-generation facilities came into operation in different laboratories across the world, amongst which was ISOLDE at CERN, still in operation today after a series of major upgrades and evolutions. The family of ion sources used throughout the facilities was enlarged over the years with new members, namely forced electron beam-induced arc discharge (FEBIAD) ion sources, surface ion sources, and later on RF-driven ion sources, more particularly electron cyclotron resonance (ECR) ion sources [3]. A schematic layout of an ISOL-type facility is shown in Fig. 2. The main components present in such facilities are:

- a primary particle driver;
- a target station with standard interfaces (vacuum, electrical power, cooling circuit) and the first optical elements of the secondary isotope beam line;
- a target and ion source unit, which is mechanically tight to the station and regularly exchanged, often by remote handling exploiting robot and piloted cranes;
- a magnetic dipole where the different components of the isotope secondary beam are selected according to mass over charge figures;
- additional electrostatic optical elements to steer the beam along the beam lines.

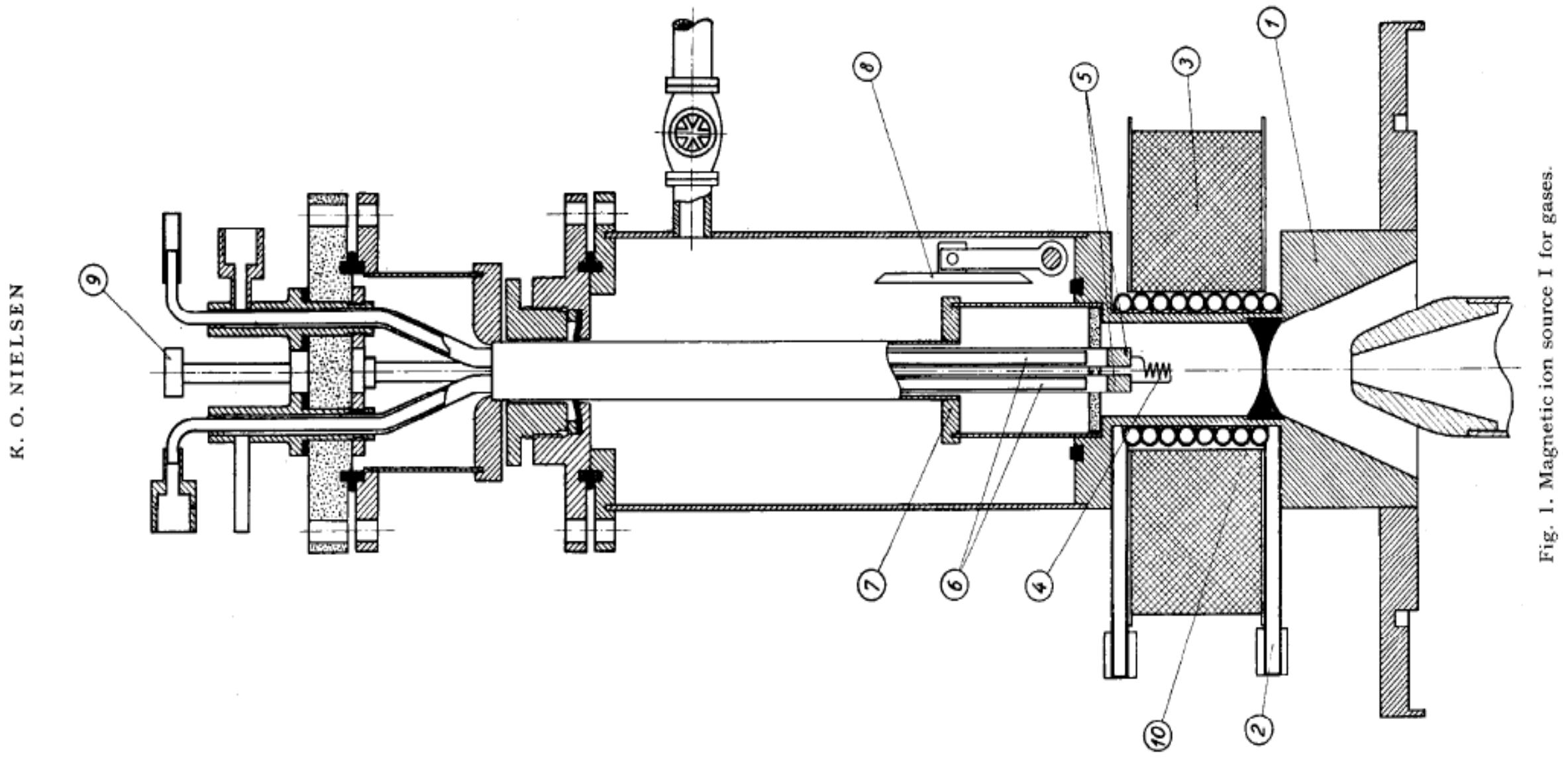

**Fig. 1:** Ion source used in the first online separation of a radioactive ion beam at the Niels Bohr Institute in 1951 [1] (reprinted from Ref. [2]). The ion source was said to be of ‘magnetic type’ by Nielsen, and is known today to be of the Nielsen type. The temperature of the arc discharge chamber could be held up to 1100°C. The 0.4 mm diameter tungsten filament lifetime was 50–100 h, operated at 0.1 mbar gas pressure. Components: 1, brass anode block; 2, Cu water cooling tube; 3, magnet coil; 4, filament; 5, Cu rods; 6, water-cooled brass tubes; 8, ‘vacuum lock’; 9, needle valve for controlling the gas inlet; 10, stainless-steel bottom plate. Operating parameters (5% efficiency on Kr): anode voltage, 100–200 V; arc current, 1–2 A; magnetic field, 0–400 G; diameter of outlet opening, 2–5 mm; beam current for krypton, 50–150 μA.

The figure of merit of a given radioactive ion beam facility comes from the folding of the following parameters:

- number of available radioactive ion beams;
- intensity of the beam;
- beam quality, for instance its purity and emittance;
- non-degradation of the beam performance over time, like intensity, purity, energy;
- facility up-time;
- (post)-accelerated beam characteristics.

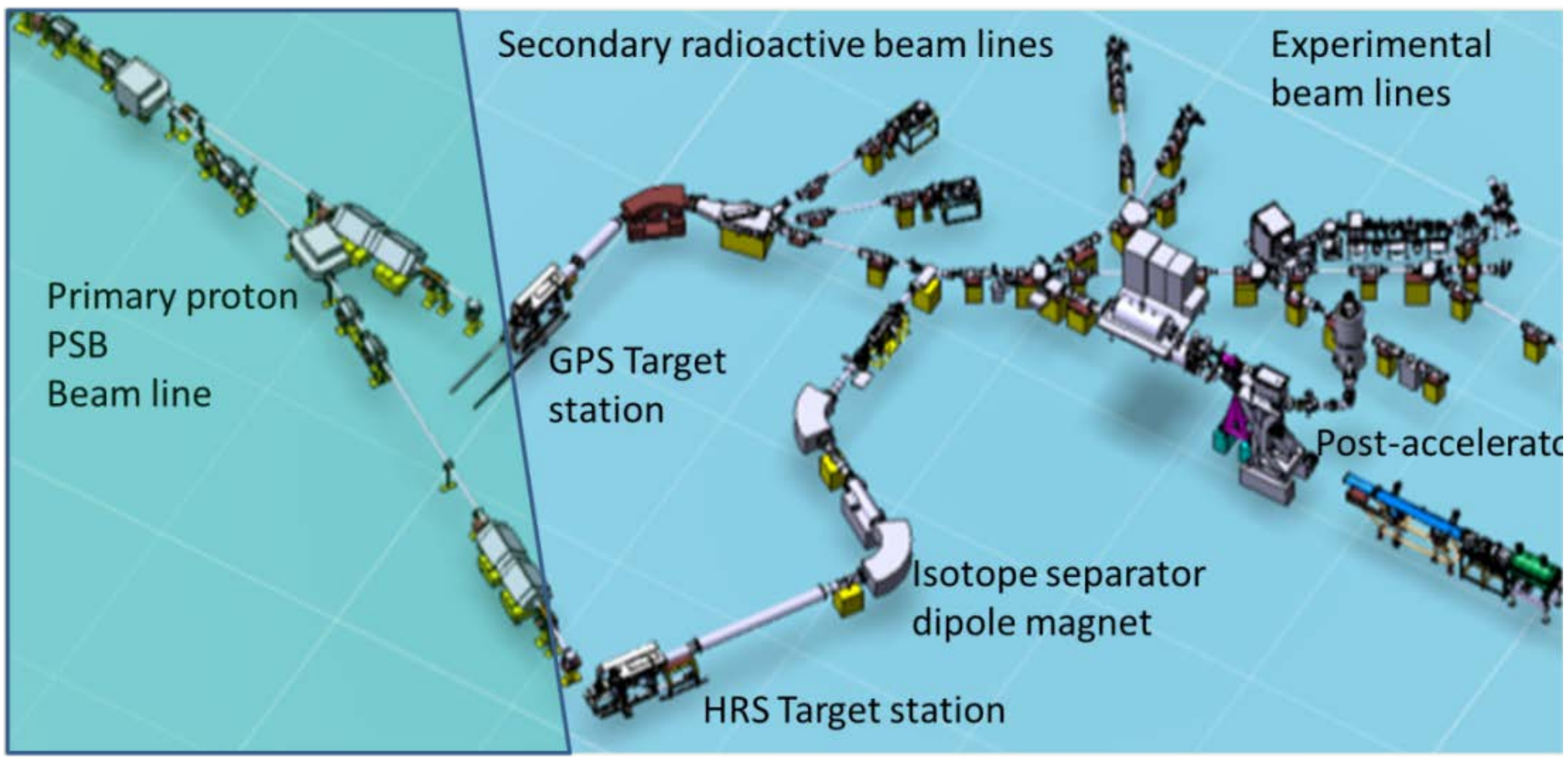

**Fig. 2:** General view of an isotope online facility such as ISOLDE at CERN. The buildings and shielding structure are not represented.

In a given facility, the characteristics of the primary beam, often a large particle accelerator, are defined in the design and construction phases. The rest of the infrastructures is also determined at the conception of the facility, such as the buildings or services. The parts that deal with the design of the target units, and to a lesser extent the beam lines, on the other hand, can still be and mostly are developed over time, to meet the evolving needs of the experiments. These developments more particularly focus on the different types of target materials to generate the isotopes, on the engineering aspects of the target production unit design, and – in the focus here – on the development of the ion sources. For a given facility, several combinations of target materials and ion sources are integrated into compact units, and are regularly exchanged in order to deliver the required isotope beams during a year of operation. The integration of the source into a compact unit dates back to the 1970s at ISOLDE (Fig. 3). A contemporary target and ion source unit is shown in Fig. 4.

## 3 Requirements for a radioactive ion source

An ideal ion source for ISOL radioactive ion beam facilities must fulfil a set of basic criteria. The development of any existing source or the design of a new one will therefore consist in satisfying the maximum number of these requirements. Often compromises have to be made, i.e., increasing the heating power of a source to achieve high ionization efficiency could affect the source stability over time.

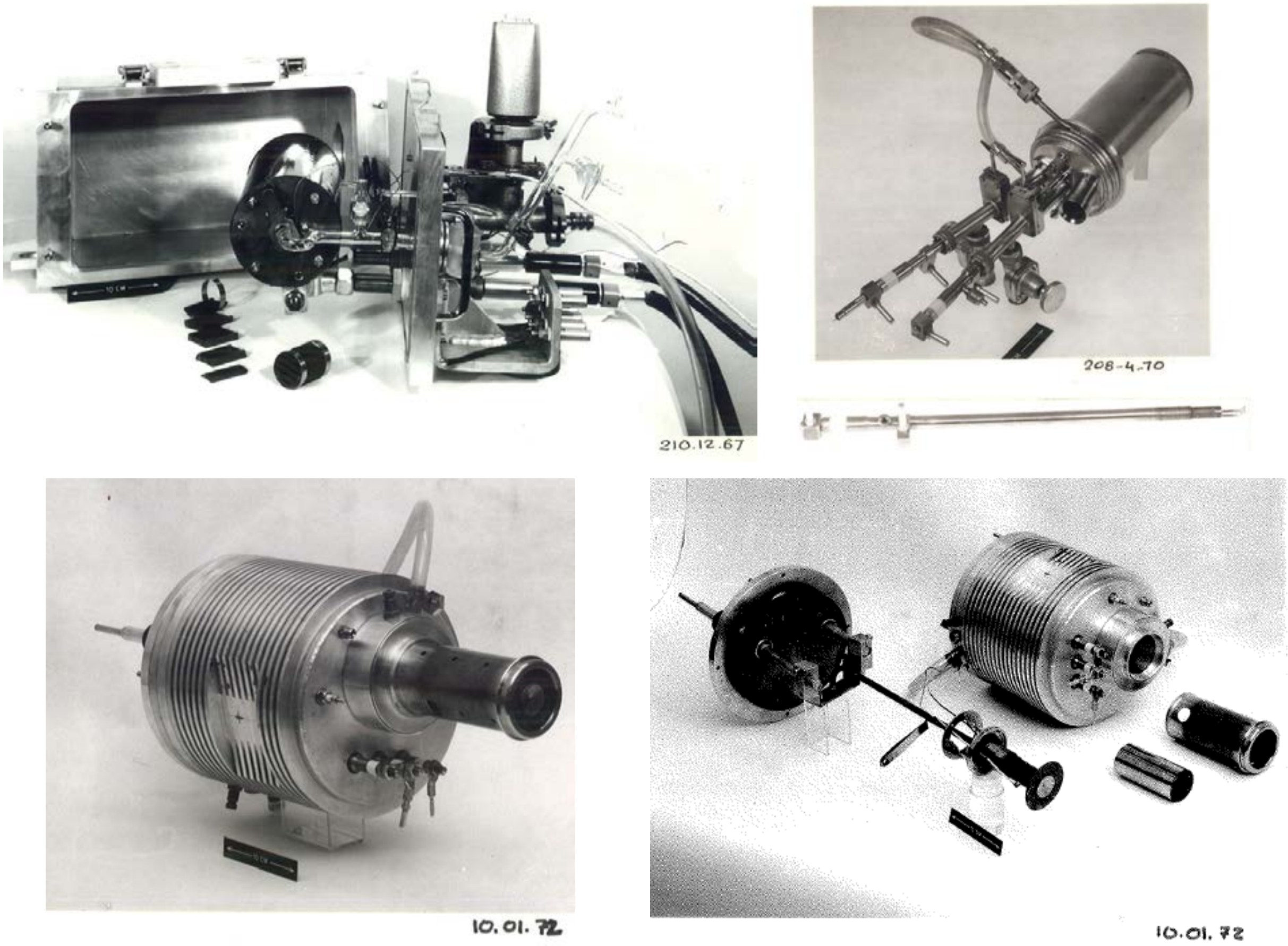

**Fig. 3:** Top: target and glow discharge ion source used at ISOLDE until the 1970s. Bottom: integrated MK-IV target and ion source unit in 1972.

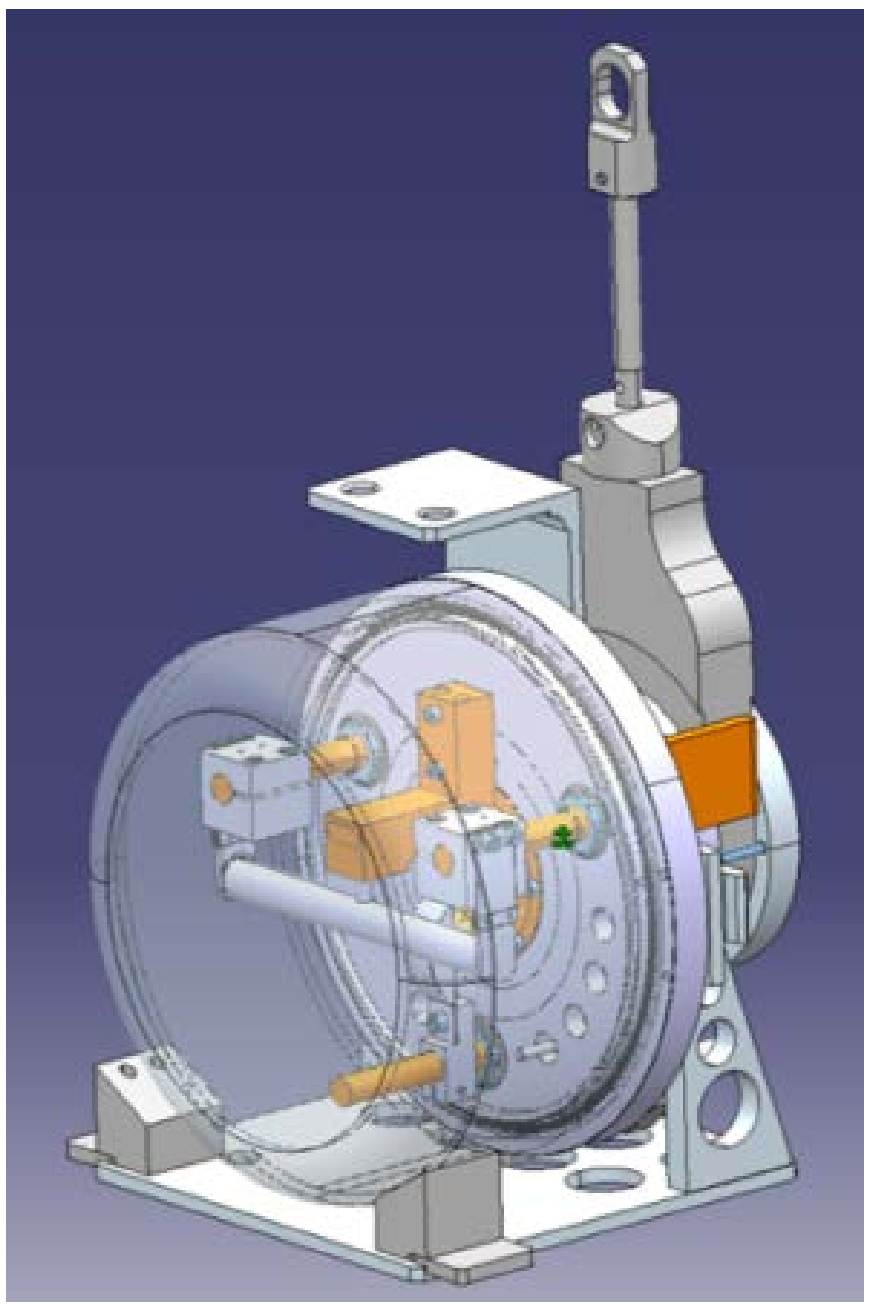

**Fig. 4:** Modern target and ion source unit at ISOLDE. The diameter of the confinement box (seen in transparency) is 30 cm.

An ideal source must:

- be compact;
- have radiation-hard components to withstand a predefined dose (e.g., 1 MGy for an installation like ISOLDE);
- be compatible with the predefined interfaces and first beam optical elements, i.e., the electrostatic extraction puller;
- provide a maximum ionization efficiency, if possible approaching 100%;
- operate in a stable manner with varying gas loads coming from the target heated at high temperatures and impacted by the primary beam (i.e., there should be no drift in performance during operation which requires a retuning of its parameters);
- be element-selective, i.e., ionize only isotopes from one element amongst a gas phase made up of other isotopes, stable impurities and molecules with several orders higher concentrations;
- produce a total extracted beam intensity compatible with the low-energy beam-line elements, i.e., typically of about 100 μA or less;
- have a perfect beam quality, i.e., no energy spread and a minimal transverse emittance;
- have small residence time with respect to the isotope of interest, i.e., in the millisecond range for the most exotic isotopes produced by the ISOL method.

It is clear that no source is universal, and since a facility aims to deliver as many different elements and isotopes as possible, sources need to be developed over the years, and, during operation, be exchanged from one production unit to the next one, typically once a week or once a month, to meet the requirements for the production of the element under consideration.

It is easy to understand that ion sources need very different characteristics to ionize noble gases such as argon, metallic elements such as nickel, or chemically reactive atoms such as boron. To help provide a more rational view, some key properties are reported in Fig. 5. A periodic table of the chemical elements is displayed, showing their ionization potential, their electron affinity and their melting and boiling points.

In the following section, the three important categories of radioactive ion sources for ISOL facilities are introduced, namely those based on surface ionization processes, on electron beam impact, and on RF-heated plasma.

## 4 Surface ion sources

### 4.1 Historical elements

Surface ion sources are used in most present-day ISOL-type facilities. They were developed in the early days of the isotope separation technique, being simple, robust, fast and selective ion sources. A layout of the target and ion source unit first developed and used in Orsay before being transferred to the Proton Synchrotron (PS) beam at CERN is shown in Fig. 6 [4]. In this configuration, the target, with direct interaction of the primary particle beam, for the production and recoil of the reaction products, is located inside an oven. It is made of graphite or refractory metal and brought up to a high temperature of 1500°C. The surfaces of the oven act as the ion source itself, exploiting the well-known surface ionization phenomenon first identified and described by Saha and Langmuir for heated metallic filaments [5]. The ions are then emitted through a vertical slit before acceleration and injection in the magnetic dipole mass spectrometer. The ionization efficiency for alkali isotope beams was reported to be small, but beams of $^{9}Li$ of 180 ms half-life could already be extracted [6]. Ten years later, the concept had been improved, and $^{11}Li$ and $^{32}Na$ beams with half-lives in the 10 ms range could be produced with a rhenium foil inserted in the oven as an ionizing surface [7]. Efficiencies in the few per cent range were reported.

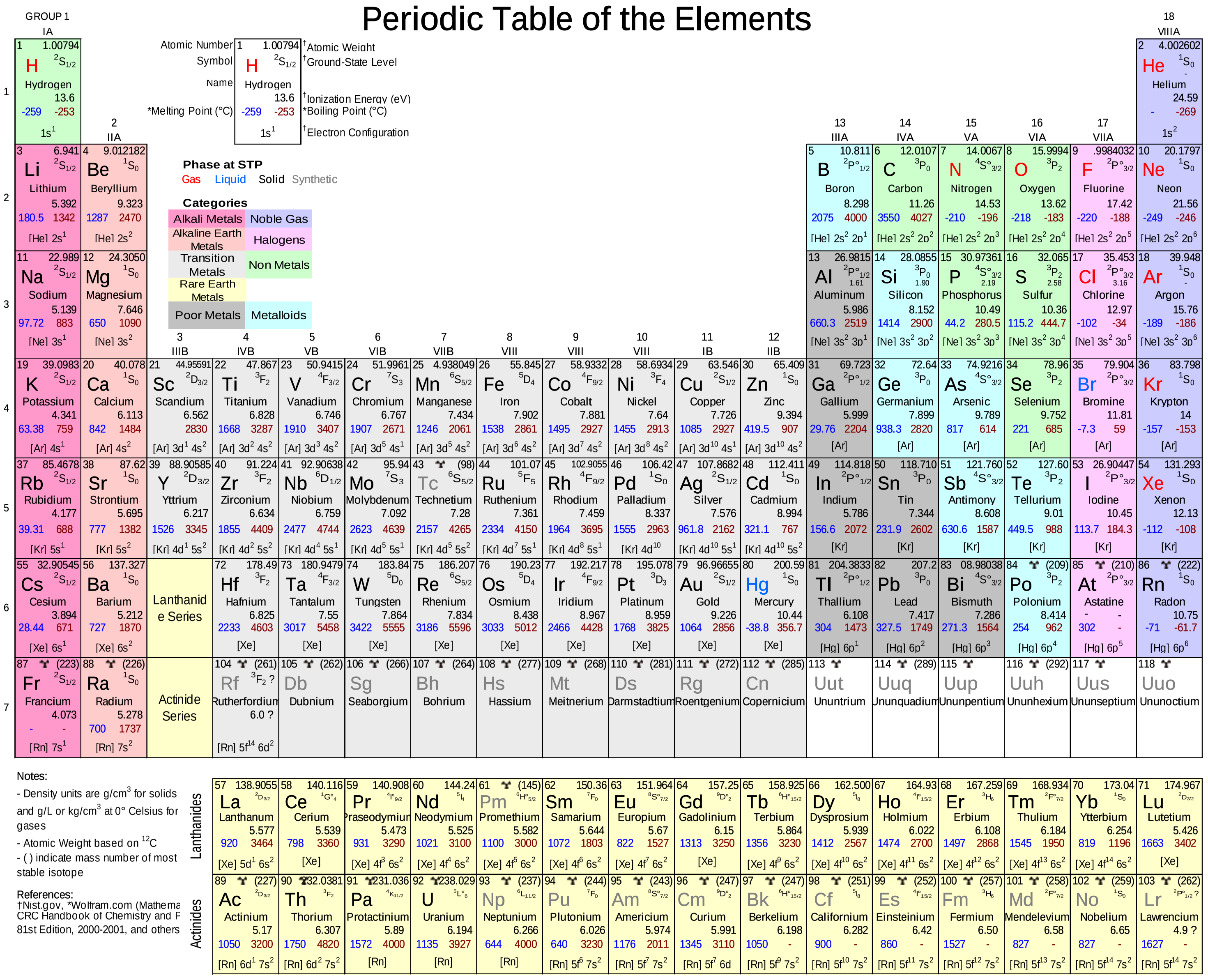

**Fig. 5**: Periodic table of the chemical elements

Surface ion sources made of hollow tubes were introduced in the 1970s [8]. They were heated by electron bombardment. A version adapted for online operation at ISOLDE was available soon after [9] (Fig. 7). In this last version, the source is connected to a large target container via a transfer line, in such a way that the beam irradiation and target heating are spatially decoupled from the operating source. Heating was provided either by electron bombardment or by ohmic current. Efficiencies in the several tens of per cent for lanthanides were reported in this configuration. Later on, other developments took place, for instance providing negative surface sources and investigating different materials [10]. They are based on similar processes as their positive counterparts for singly negative ions. Some examples of such sources are shown in Fig. 8.

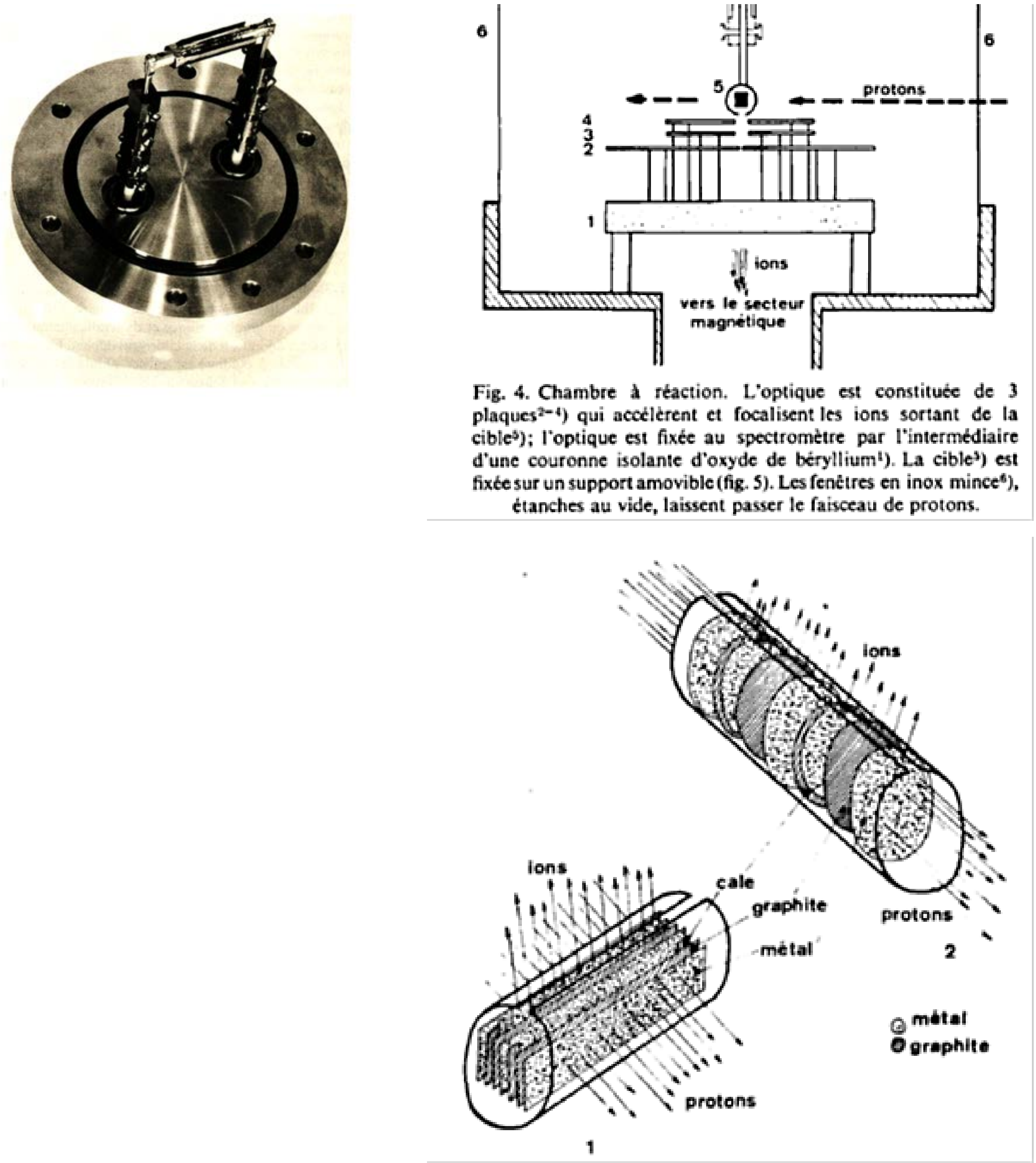

**Fig. 6:** Surface ion source developed at Orsay to ionize alkali metal beams and used at CERN PS in the 1960s (from Ref. [4]).

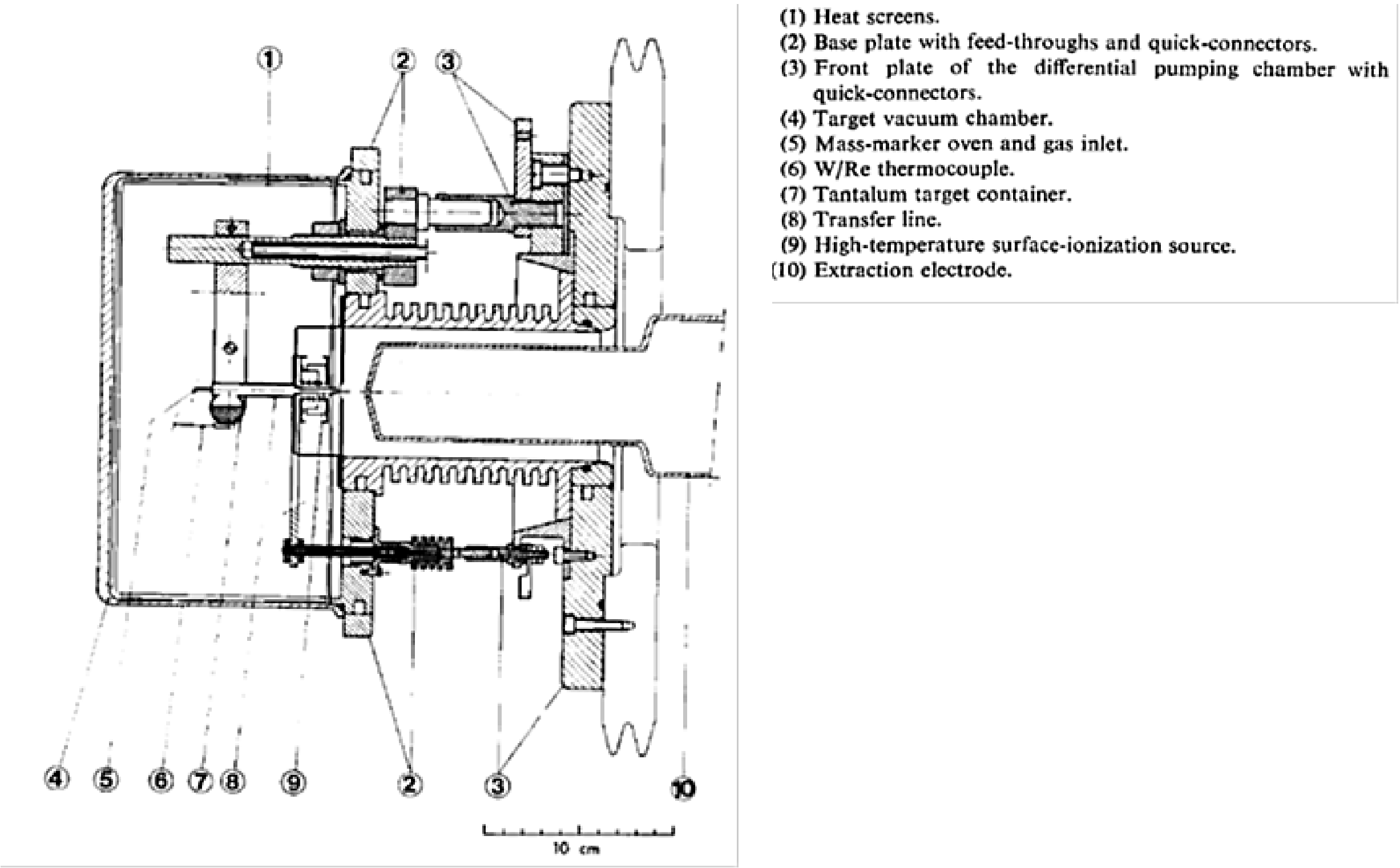

**Fig. 7:** Surface ion source used at CERN-ISOLDE (from Ref. [9])

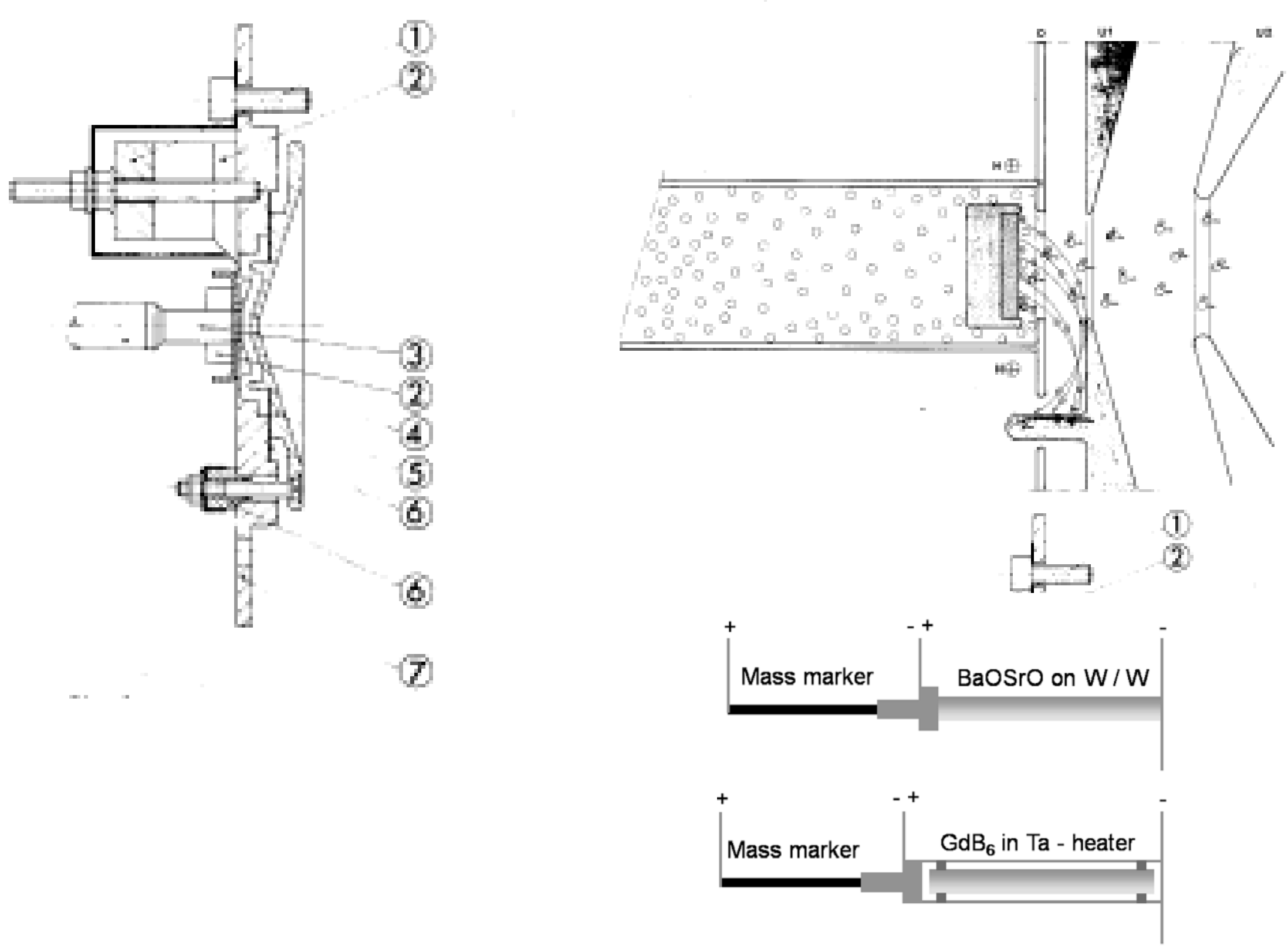

**Fig. 8:** MK4 negative surface ion source developed at ISOLDE and some prototype tubular geometries (from Ref. [10]).

### 4.2 Mode of operation and underlying physics principles

The principle of operation is rather simple. It is based on the following well-known Saha–Langmuir equation, which describes the ionization degree $\alpha$ and efficiency $\varepsilon$ for weak plasmas [5] and for heated surfaces, where $g_+$ and $g_0$ are the quantum degeneracy from the charged and neutral states, $W$ is the work function of the ionizing material, and $\phi_i$ is the ionization potential of the atoms under consideration:

$$\alpha_{\text{surface}} = (g_+/g_0)\exp[(W-\phi_i)/kT]\,, \tag{1}$$

$$\alpha_{\text{surface}} = \exp[(W-\phi_i^*)/kT] \qquad \text{with} \qquad \phi_i^* = \phi_i - kT\ln(g_+/g_0)\,, \tag{2}$$

$$\varepsilon_{\text{surface}} = \frac{1}{1+1/\alpha_{\text{surface}}}\,. \tag{3}$$

Eqs. (1) and (3) were established in 1925 and were used to describe the measured efficiencies of ion sources of simple geometries, such as heated surfaces and filaments. Eq. (2) is a convenient form in which the pre-factor originating from multiple quantum states of the neutral atom and the charged ion are already included in a normalized ionization potential parameter $\phi_i^*$.

However, this model was found to be inaccurate in properly describing the behaviour of the source in a hollow-tube configuration. Corrections to account for the proper plasma parameters in such heated tubes or capillaries had to be included.

It took ten years to develop a satisfactory model and to test it against experimental data. More details can be found in two review articles written by R. Kirchner [11]. In a hollow-tube configuration, correction factors are required to properly describe the observed ionization efficiencies. In this case, a very low-density plasma is created in the tube; this introduces a first correction factor $\kappa$, which accounts for the average number of collisions that the isotope experiences with the hot surface of the tube before being extracted, and a second one $\omega$, the trapping efficiency of the emitted ion inside the plasma potential well. Some arguments based on the geometry and plasma properties can be given to estimate these two factors; however, in practice, the product $\omega\kappa$ is used as a free parameter to fit a set of experimental data obtained with a series of chemical elements that span a relevant range of ionization potentials and efficiencies. The model in Eq. (4) is therefore now accepted to describe ionization efficiencies for hollow-tube ionizers as found in most of the present ISOL-type facilities:

$$\varepsilon_{\text{ionizer}} = \frac{1}{1+1/\omega\kappa\alpha_{\text{surface}}}\,. \tag{4}$$

A graph plotting $\varepsilon_{\text{ionizer}}$ versus $\phi_i^*$ for a source made of tungsten with two different partial pressures of neutral xenon gas is shown in Fig. 9. A similar graph has been obtained very recently at JAEA-ARSC with a hot tantalum cavity as shown in Fig. 10 [12].

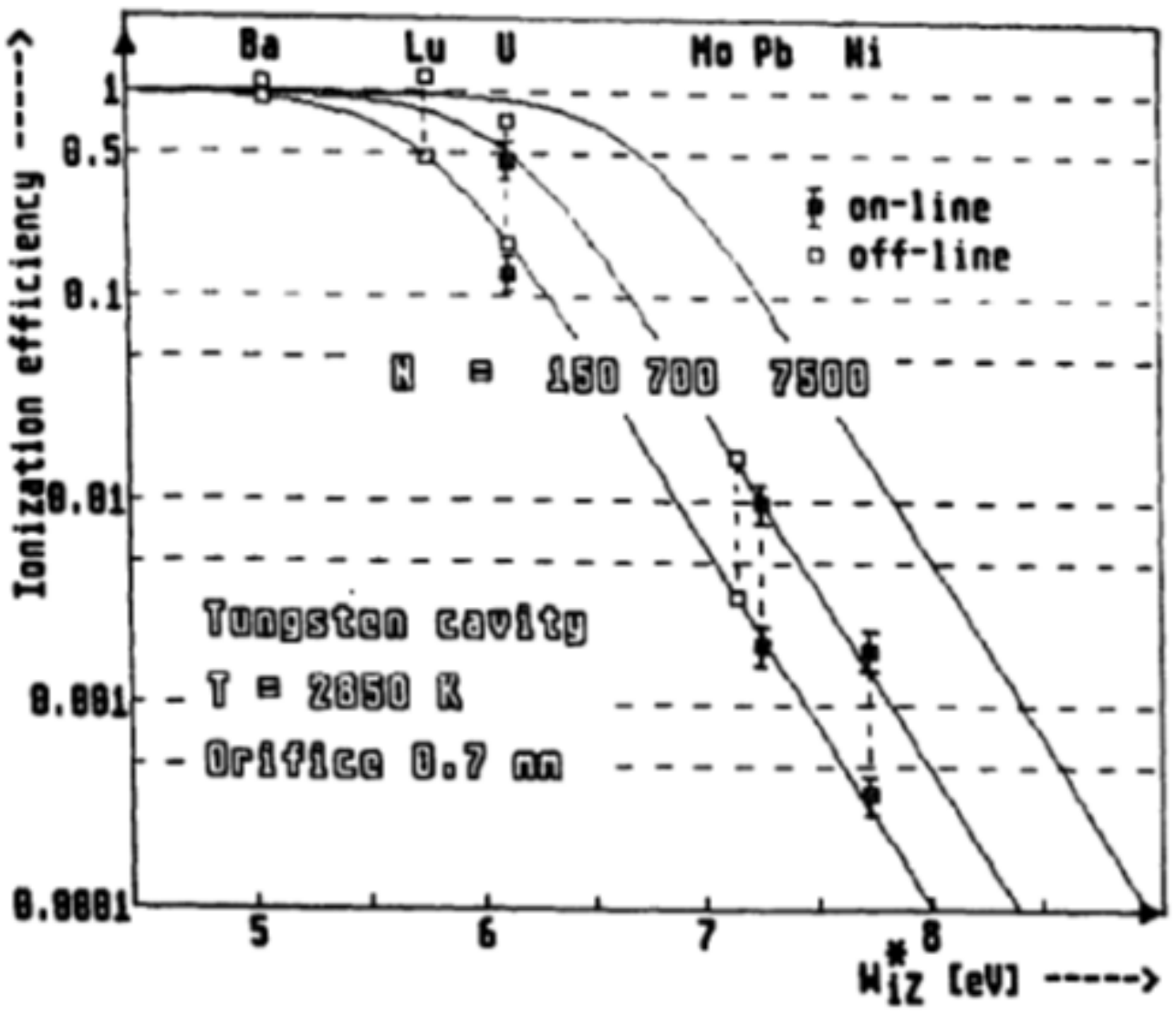

Fig. 5. Ionization efficiency of various trace elements in a cavity operated without xenon (lower values) and with a xenon input of around $10^{-3}$ $cm^3$(STP)/s (upper values), respectively. The latter corresponds to a neutral density of $\approx 5 \times 10^{14}/cm^3$ in the cavity. Curves are calculated according to eqs. (7) or (9). $N = 700$ implying that all surface-ionized particles leave the cavity as ions.

**Fig. 9:** Efficiency plotted versus ionization potential of different condensable chemical elements. Two sets of data recorded at different partial pressure in the hollow cavities are reported. The fit of the data using Eq. (4) and different values for $\omega\kappa$ (referred to as $N$ in the figure) are also shown. From Ref. [11].

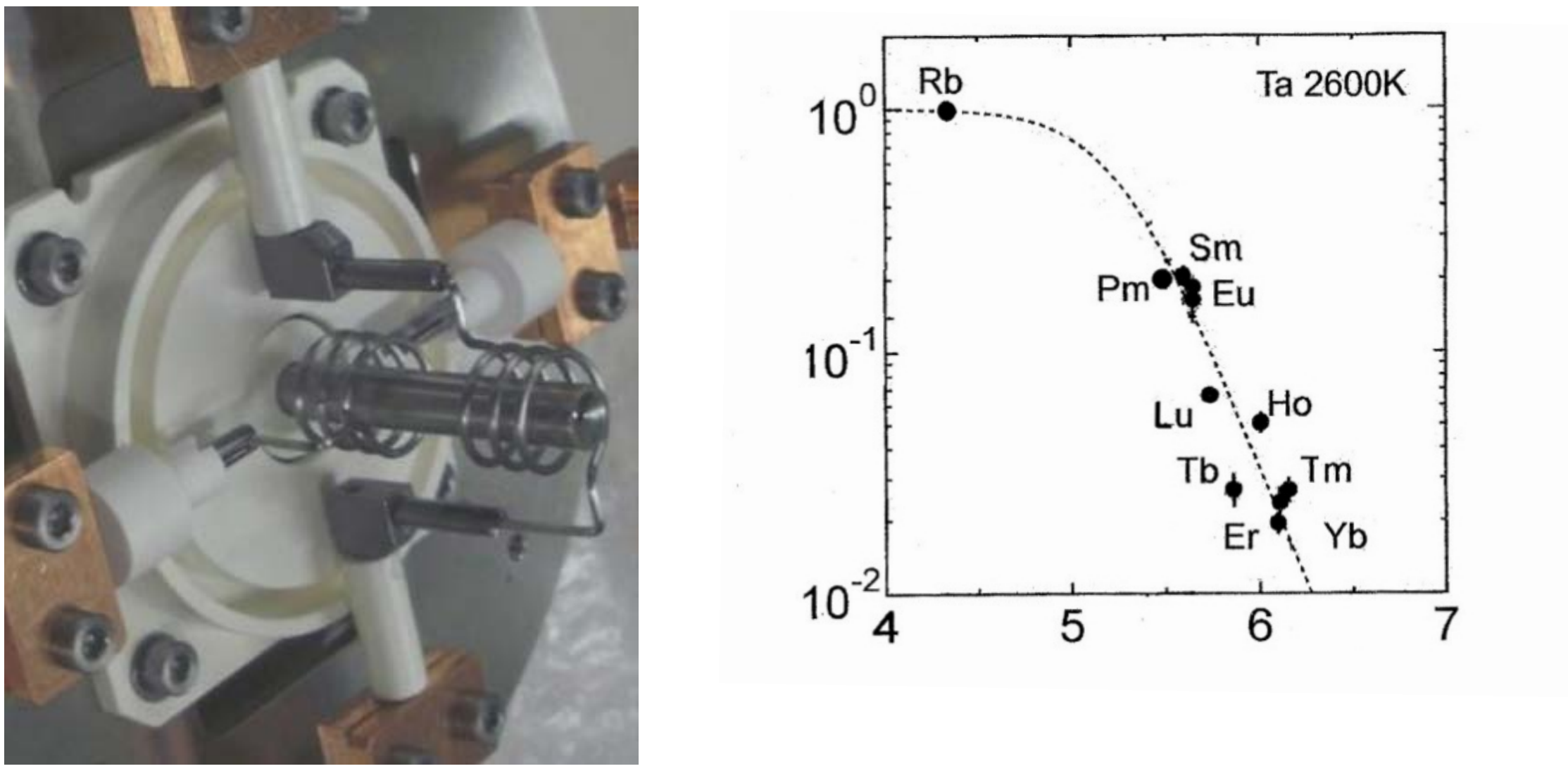

**Fig. 10:** Tantalum surface ion source developed at ASRC-JAEA heated by electron bombardment with two filaments. The ionization efficiencies are reported for a set of exotic radio-lanthanides for a cavity operated at 2600 K [12].

### 4.3 Specific applications, future developments

Surface ion sources are used in most ISOL-type facilities in operation or in construction. They are exploited to produce the most intense alkali and alkaline-earth isotope beams, and provide in addition some chemical selectivity. Besides their utilization as selective ionizers, they are nowadays combined with RILIS lasers to ionize metallic elements. This last mode of ionization is described elsewhere in this present volume by B. Marsh [13]. In this case, choices of materials with lower work functions are generally preferred, with a successful demonstration of using ceramics of $GdB_6$ to suppress isobaric contaminants [10]. Surface ion sources have been used to produce molecular beams with suitable ionization potentials. This is mainly lanthanide oxide and fluorine compounds, and alkaline-earth fluoride molecules [9, 14]. In that case the theoretical treatment to account for the operational behaviour of the source has not been investigated in great detail. The main elements required to account for evolving intensities and efficiencies of atomic versus molecular ions have been reported in the case of Ba and O atoms for $Ba^+$ and $BaO^+$ ion production [15]. The interested reader should refer to that publication for more details.

## 5 FEBIAD ion sources

### 5.1 Historical elements

Forced electron beam-induced arc discharge (FEBIAD) ion sources were developed to overcome some of the drawbacks observed while operating the previous generation of sources equipped with a heated cathode emitting electrons in a high-pressure plasma chamber. In that previous generation, the configuration of the filament, the magnetic field, the polarization of the plasma chamber walls, i.e., anode, and the layout of the ion extraction led to different versions referred to as the Nielsen, Bernas–Nier or hollow-cathode ion sources [11].

One version of the Nielsen source was shown in Fig. 1 and another one is shown in Fig. 11. Reported drawbacks are a finite lifetime of the cathodes, which could be dependent on the chemical reactivity of the gas phases, operation under high pressures of $10^{-2}$ mbar, high total extracted beam intensities, and difficulties in reaching and maintaining stable modes of operation.

The principal technical modifications of the previous source designs were the change of the cathode, from coiled wire geometry to a flat disc, and the insertion of a polarized grid in front to stabilize the electron emission current. The first development and application of these sources were reported by R. Kirchner at GSI and S. Sundell at Isolde in the 1970s [16]. Different versions of the FEBIAD sources were developed, exploiting various structural materials for the cathode, the accelerating grid, the anode chamber or the heat screens. Figure 12 shows the three FEBIAD sources MK3, MK5 and MK7 out of a larger series developed at ISOLDE and used until 2009 [17]. The different models were combined with different types of targets and connected via transfer lines kept at different temperatures.

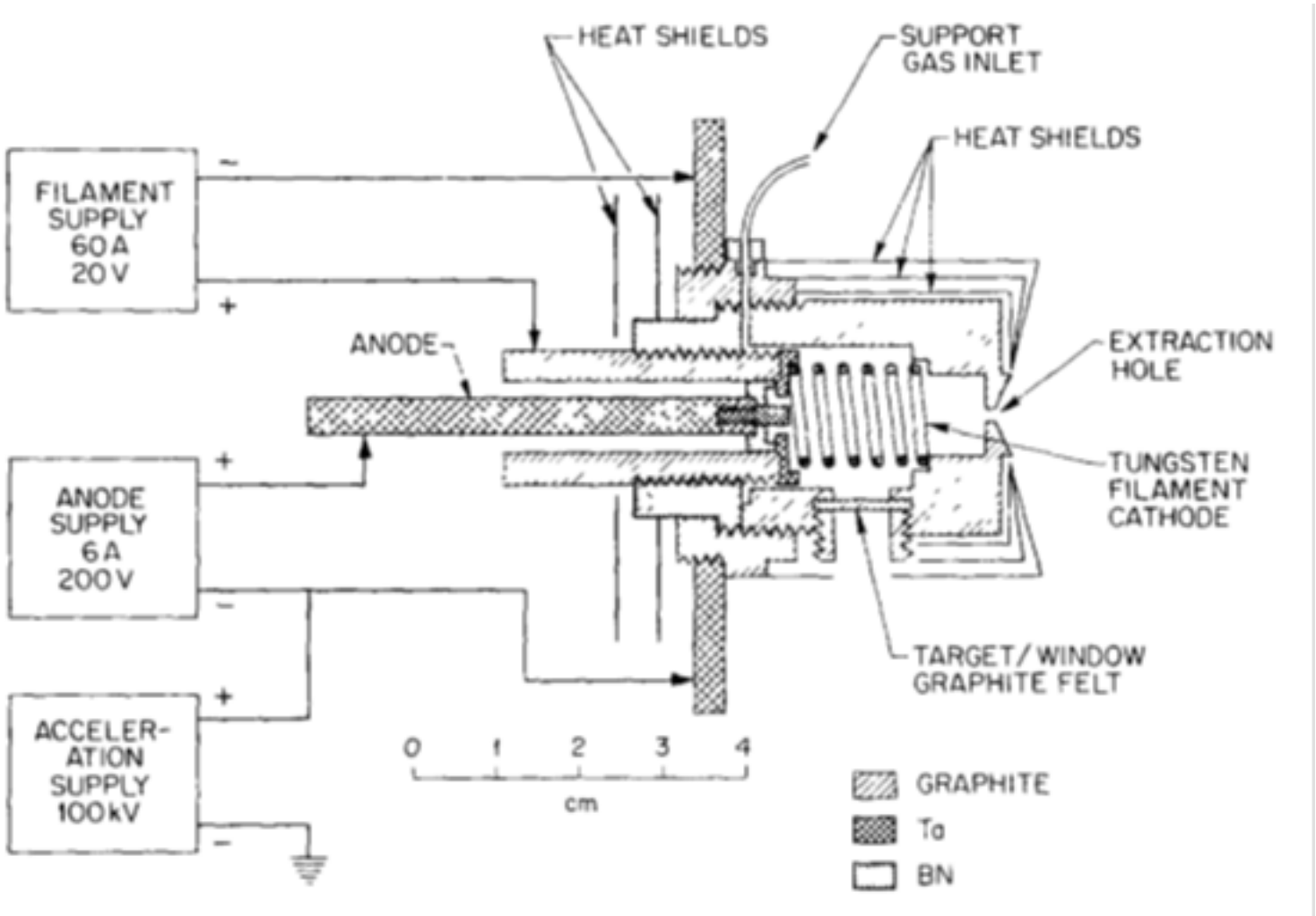

Fig 2 Hollow cathode ion source, UNISOR [19]

**Fig. 11:** Hollow-cathode ion sources applied at UNISOR at GSI (from Ref. [11])

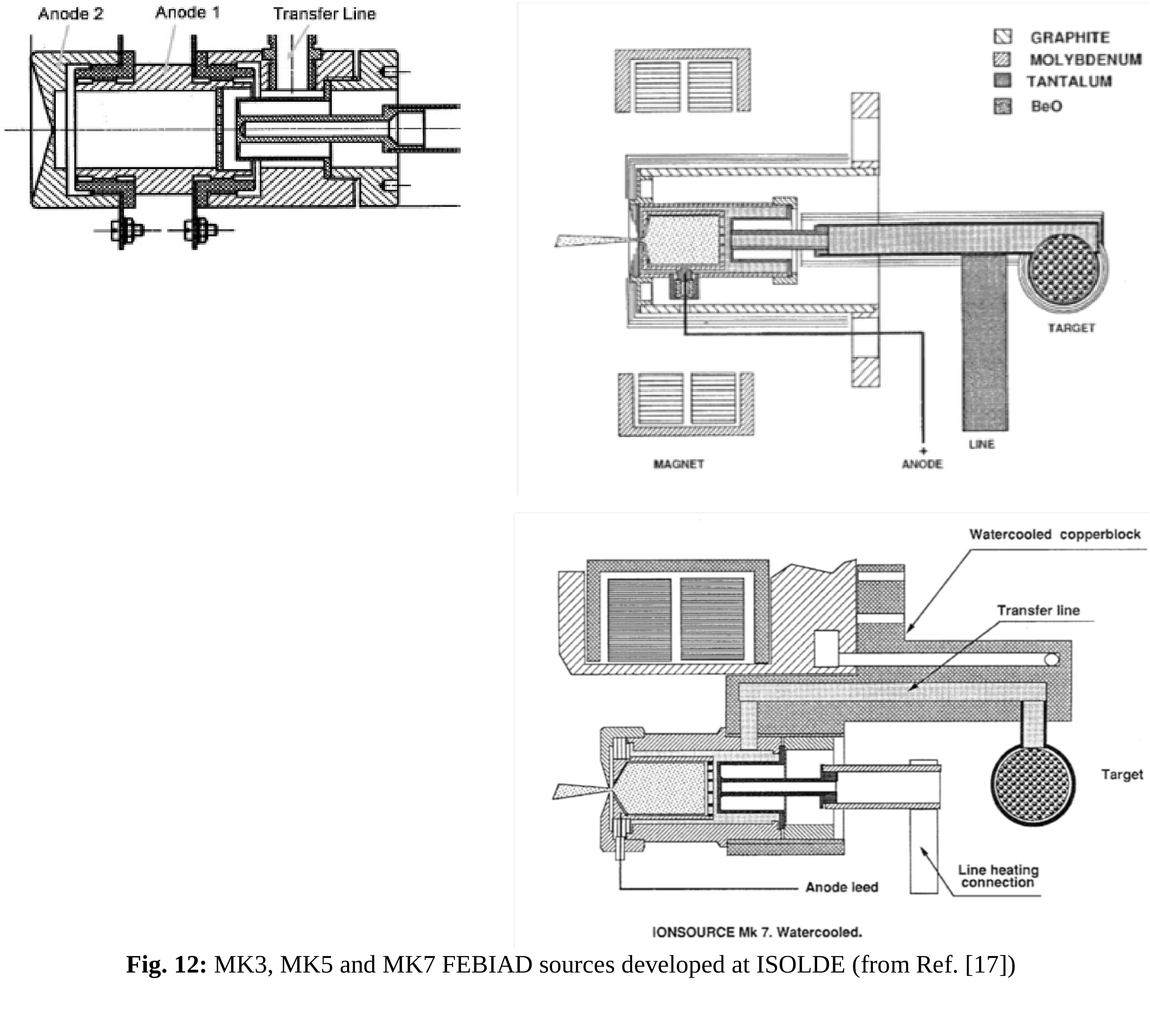

**Fig. 12:** MK3, MK5 and MK7 FEBIAD sources developed at ISOLDE (from Ref. [17])

### 5.2 Mode of operation and underlying physics principles

None of the theoretical models put forward could well describe the reported experimental efficiencies of the FEBIAD sources over extended ranges of operational parameters. In addition, some of the used parameters were not physically sound. During his PhD, L. Penescu proposed a new model that seemed to provide a reasonably good description of the observed ion beam currents using simple physical parameters [18]. This model and detailed experimental investigation of the three FEBIAD sources in operation at ISOLDE led to a new generation of such sources, which we named VADIS for versatile arc discharge ion source.

The proposed ionization efficiency model for FEBIAD sources is the following:

$$\varepsilon = \frac{R_{\text{ion}} V_{\text{source}} f_{\text{extr}}}{I_{\text{n_in}}} = \frac{(n_{\text{e}} n_{\text{n}} \sigma_{\text{ion}} v_{\text{rel}}) V_{\text{source}} f_{\text{extr}}}{I_{\text{n_in}}} \,. \tag{5}$$

The model described by Eq. (5) is a linear combination of a set of parameters that can be calculated from an experimentally accessible set of data, such as the emitted electron current from the cathode or the injected neutral buffer gas from the calibrated leak rate. Indeed, the different terms in Eq. (5) can be explicitly computed:

$$v_{\text{rel}} \approx v_{\text{e}} = \sqrt{\frac{2eU}{m_{\text{e}}}} \tag{6}$$

is the average relative speed between the neutral isotopes and the electrons emitted from the cathode and accelerated by the anode grid at a potential $U$, typically at 30–250 V;

$$n_{\text{e}} = \varepsilon_{\text{e}} \frac{j_{\text{e}}}{v_{\text{e}}} \,, \tag{7}$$

$$j_{\text{e}} \ [\text{mA mm}^{-2}] = \frac{4}{9} \varepsilon_0 \left( \frac{2e}{m_{\text{e}}} \right)^{1/2} \frac{U^{2/3}}{d^2} \,, \tag{8}$$

with $n_{\text{e}}$ the electron density in the source, $\varepsilon_{\text{e}}$ a scaling factor that accounts for the grid transparency and electron reaching the drift plasma chamber, and $j_{\text{e}}$ the electron current emitted from the cathode, described by the Child–Langmuir equation (8) when space charged limited, with $U$ the acceleration potential of the grid and $d$ the distance between the cathode and the grid. This is for sources operated typically at temperatures above 2000°C in the configuration of a VADIS source. For low temperatures, the electron emission is then limited by the Richardson–Dushmann electron emission from a cathode:

$$j_{\text{e}} \ [\text{mA mm}^{-2}] = AT^2 \exp(-W/kT) \qquad \text{with} \qquad A \cong 1200 \ \text{mA mm}^{-2} \ \text{K}^{-2} \,. \tag{9}$$

Here the work function $W$ of the cathode is in eV and $A$ is the Dushman constant, which can adopt different values for different materials. In Fig. 13 the current measured at the anode power supply is recorded, and can as a first approximation be compared to Eqs. (8) and (9), using the recorded temperature of the cathode and the anode potential. The neutral atom density is expressed by Eq. (10), in which the gas leak rate $Q_{\text{in}}$ has been measured with a tracer gas, $M$ is the mass of the isotope, the temperature $T$ is assumed homogeneous and is that of the cathode, and $S_{\text{out}}$ is the surface area of the extraction hole of the source:

$$n_{\mathrm{n}} = 2\times10^{18} Q_{\mathrm{in}} \frac{\sqrt{M}}{T^{3/2} S_{\mathrm{out}}}. \tag{10}$$

The 1+ ionization cross-section is provided by the Lotz formula:

$$\sigma_{+} = 4.5\times10^{-14} \sum_{nl} \frac{\ln(E_{\mathrm{e}}/E_{1+,nl})}{E_{\mathrm{e}} E_{1+,nl}} \ [\mathrm{cm}^2], \tag{11}$$

And finally $V_{\mathrm{source}}$ is the volume of the source, $I_{\mathrm{n_in}}$ is the isotope current entering the source, and $f_{\mathrm{extr}}$ is the extraction efficiency once an ion is produced.

By folding together Eqs. (5)–(11), the ionization efficiency can be expressed as

$$\varepsilon = 2.33\times10^{4} f V_{\mathrm{source}} A \exp(-W/kT) l \frac{\ln(U/\phi_{\mathrm{i}})}{U\phi_{\mathrm{i}}} \frac{\sqrt{M}\sqrt{T}}{S_{\mathrm{out}}}. \tag{12}$$

Equation (12) can be tested for different types of FEBIAD ion sources, and different parameters of operation. Values of $V_{\mathrm{source}}$ and $S_{\mathrm{out}}$ are directly deduced from the mechanical design. $W$ is a known property of the cathode (classically made of graphite or refractory metals), provided the surface is free of contaminants. The temperature of the source $T$ is deduced from a calibration curve reporting $T$ measured at the surface of the cathode, with respect to the applied electrical heating power. $M$ and $\phi_{\mathrm{i}}$ are the mass and ionization potential of the isotope under consideration. $U$ is the anode potential which sets the energy of the electrons. The last parameter $f$ cannot easily be independently computed. However, it directly reflects if the design of the source and its mode of operation are chosen to optimize the electron–ion interactions and ion extraction before losses on the walls of the chamber. It is typically affected by the grid transparencies for the emitted electrons, by the internal pressure, by the configuration of the extraction geometry, by the penetration of the electrostatic field of the beam extraction electrode and by the magnetic field deflecting the electron trajectories. The efficiencies are plotted in Figs. 13 and 14 for a VADIS source, with the characteristic exponential increase in a first part, the saturation in a second region, and an eventual decrease in a last part.

As can be seen in Fig. 13, above about 2000°C, the electron emission of the cathode becomes space-charged-limited and follows the Child–Langmuir Eq. (8). In this case, Eq. (12) can be re-expressed accounting for the proper formulation of the electron current $j_{\mathrm{e}}$:

$$\varepsilon = 5.42\times10^{-2} f V_{\mathrm{source}} \frac{U^{3/2}}{d^2} l \frac{\ln(U/\phi_{\mathrm{i}})}{U\phi_{\mathrm{i}}} \frac{\sqrt{M}}{S_{\mathrm{out}} T^{3/2}}. \tag{13}$$

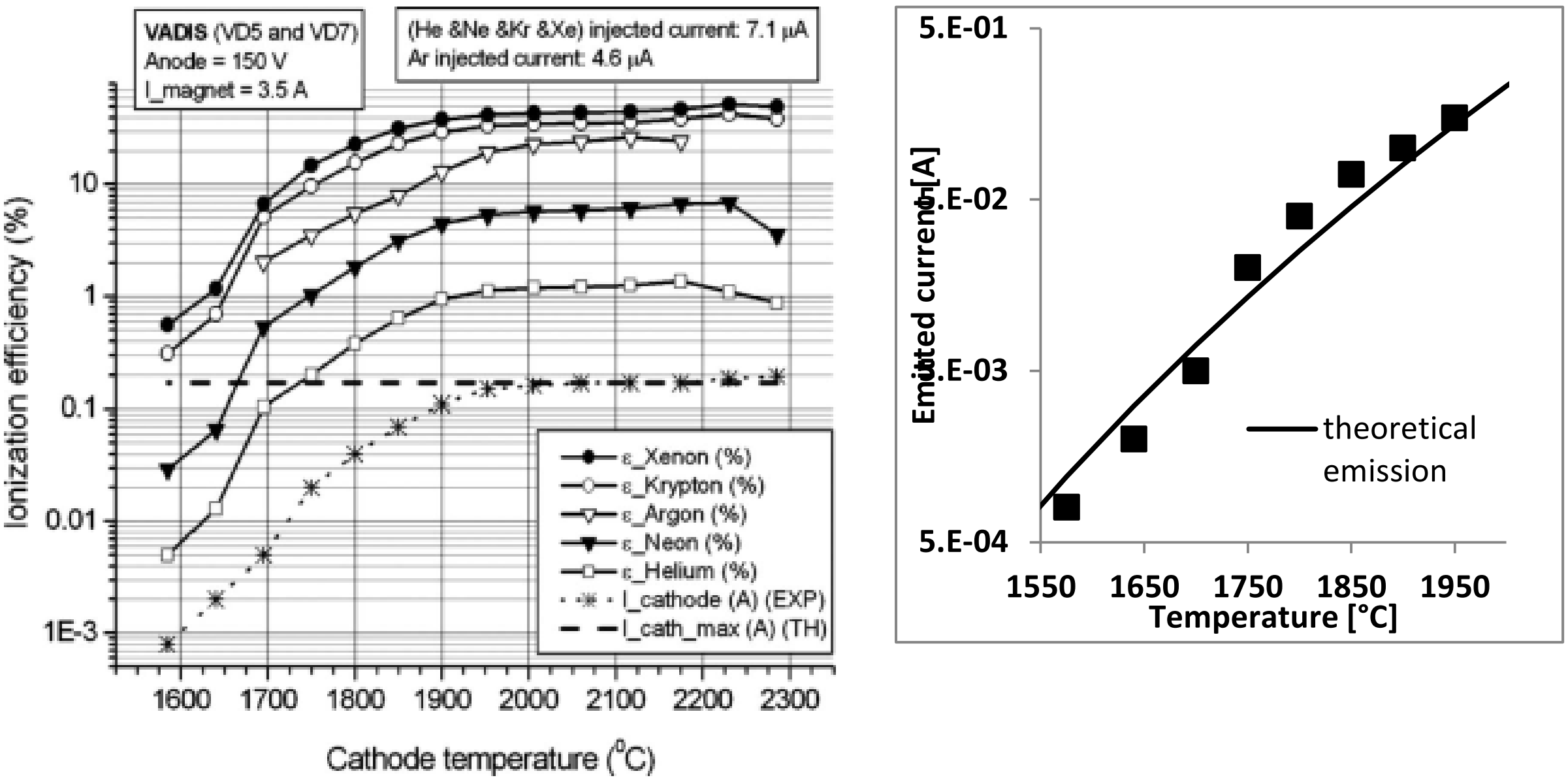

**Fig. 13:** Left: efficiencies measured for a series of noble gases in a VADIS source, electron emission from the cathode, and theoretical limitation predicted from Child–Langmuir space-charged emission (from Ref. [18]). Right: electron emission superimposed upon the theoretical prediction of Eq. (9), with $A$ = 600 mA mm$^{-2}$ K$^{-2}$, $W$ = 4.12 eV and $S_{\text{cathode}}$ = 1.13 cm$^2$.

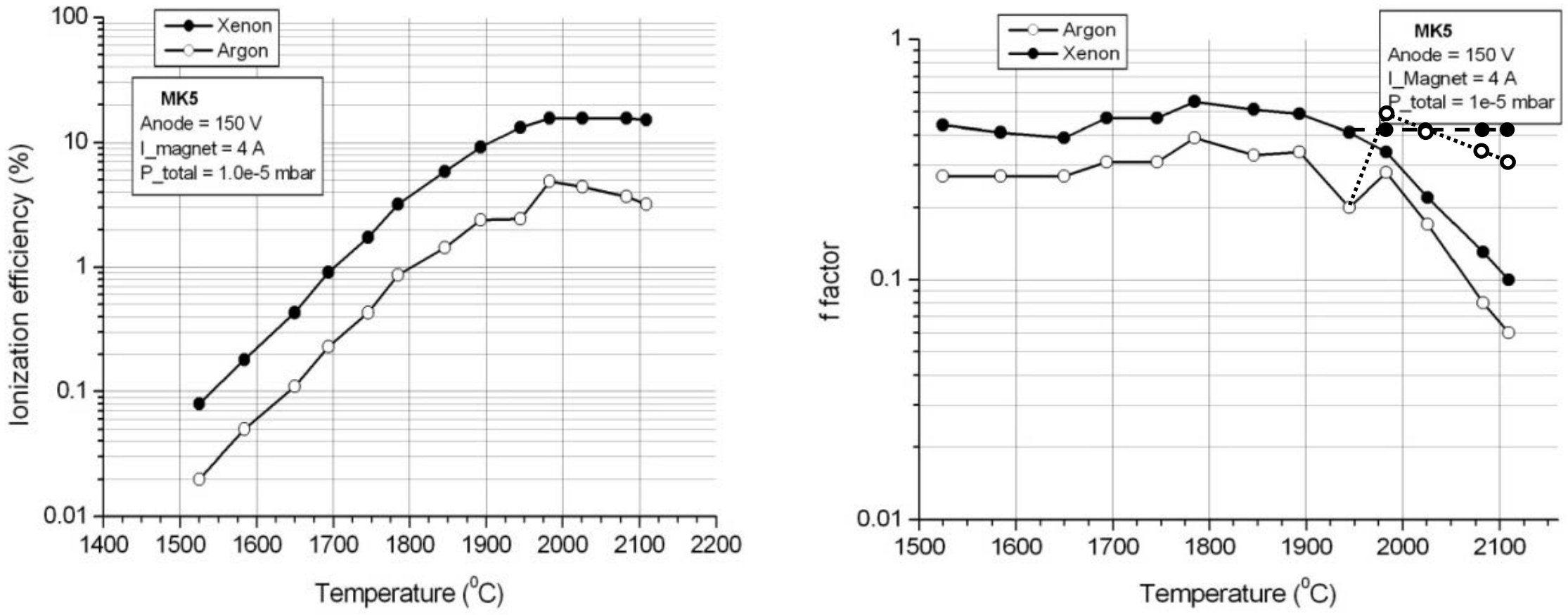

**Fig. 14:** Efficiencies measured in a MK5 source, and $f$ factor deduced from Eq. (12). The dashed lines are obtained when the cathode emission is space-charge-limited, Eq. (13) [18].

The deduced $f$ factors are also shown in Fig. 14. When the electron emission is properly measured or computed with appropriate models, this factor gives important insights into the quality of the design, and an optimal range of operational parameters. In particular, it shows if the design of the source can be improved and what are the elements that need modification.

## 5.3 Specific applications, future developments

The FEBIAD sources are versatile sources, compact, fast, and provide high ionization efficiencies for condensable elements. They have also been used to produce molecular beams of condensable elements such as $SnS^+$, $SeCO^+$ and $AlF^+$ [19]. They can be adapted to match specific requirements such as an improved beam quality from a reduced extraction hole surface area, a reduction of $CO_2^+$ beam impurities replacing graphite parts, and so forth. Interesting developments are the on-going coupling of these cavities with laser sources, and should provide exciting developments in the near future.

# 6 RF heated sources

## 6.1 Historical elements

This last section covers radio-frequency-heated sources applied for radioactive ion beams. This type of source was applied for radioactive ion beam production much more recently than the other two preceding classes. For instance, ECR sources, extensively described in these series of lectures for stable beams, have been proposed and developed for operation in different facilities. Such sources were developed for stable beams in the 1970s. They were applied for radioactive beam generation about 20 years after, at GANIL in France, in Louvain-La-Neuve in Belgium and at the TISOL facility in Canada [20, 21]. These sources are mostly used to ionize molecular beams that are volatile at room temperature. The source operated in Louvain-La-Neuve to produce intense beams of oxygen and nitrogen isotopes is shown in Fig. 15. This type of source, however, presents a number of challenges for operation in the harsh environments found at target stations of the ISOL facilities. As can be seen in Fig. 15, this was avoided by positioning the source in a remote position with respect to the target position. However, this is at the expenses of the speed of transport of the radioactive isotopes from the target to the ion source. The combined graphite target and Nanogan 3 ion source is shown in Fig. 16.

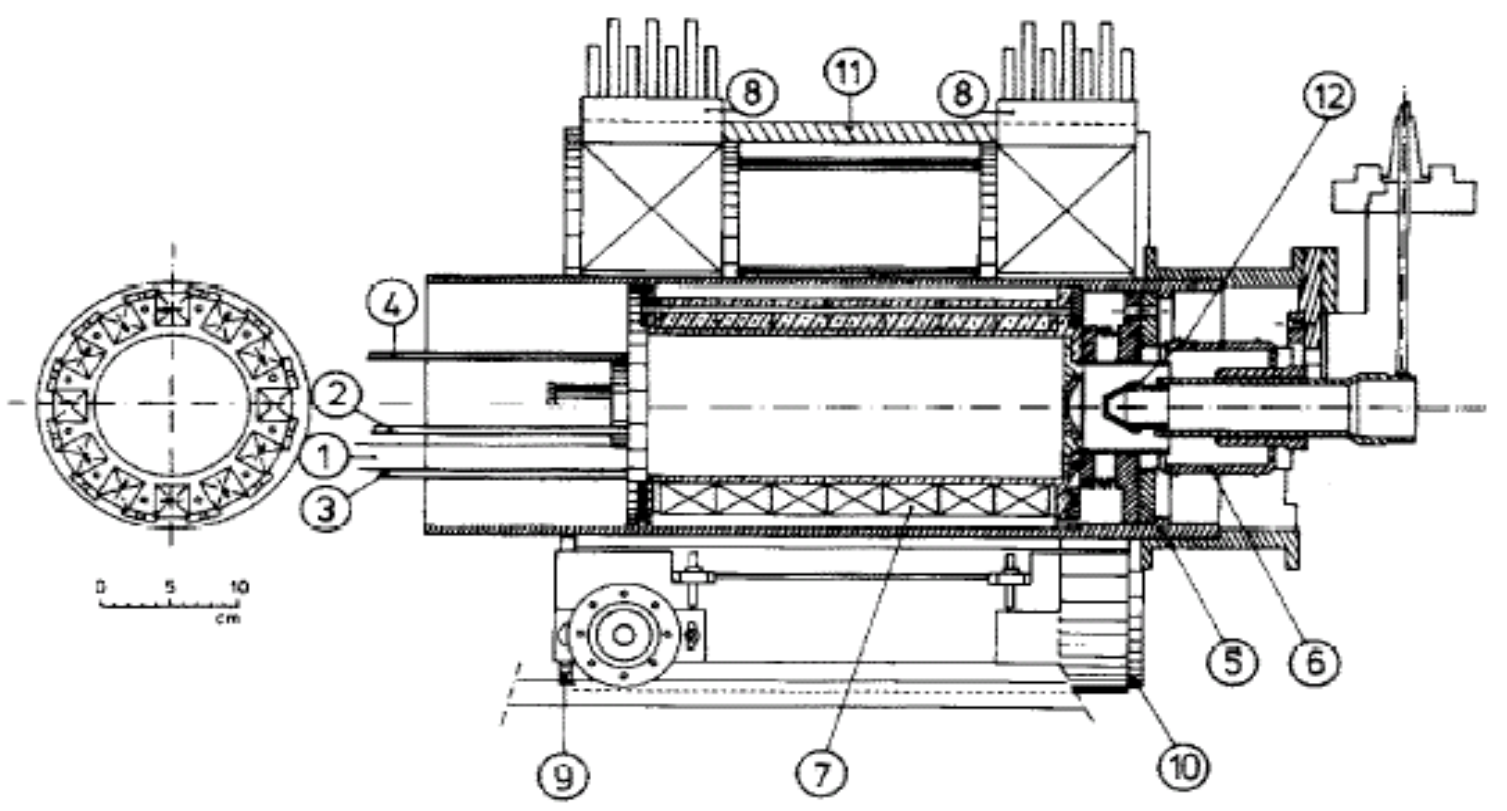

**Diagram of the ECR ion source:** micro wave guide (1), cooling water in/outlet (2, 3), gas input (4), HV insulator (5, 6), permanent magnetic bars (7), solenoids (8), iron yoke (9, 10, 11), extraction electrode (12).

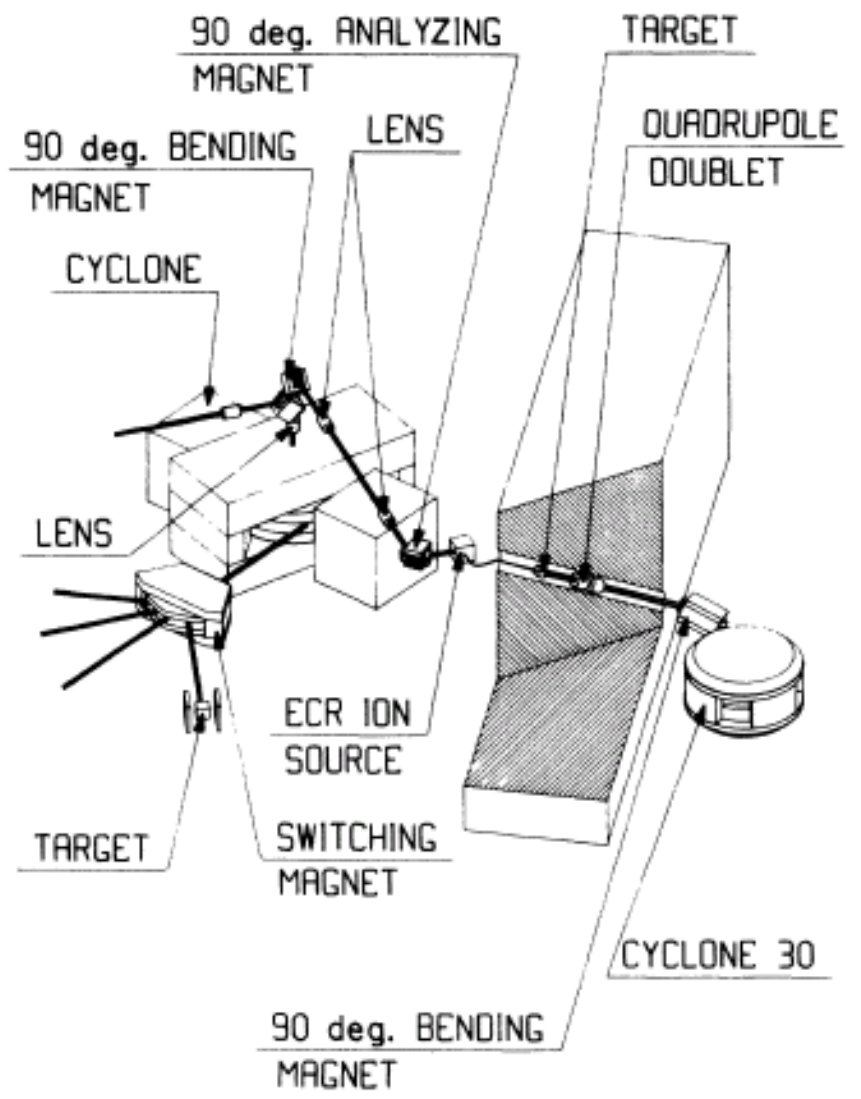

**Fig. 15:** Top: ECR source operated at CRC, Louvain-La-Neuve (from Ref. [20]). Bottom: layout of the facility [21].

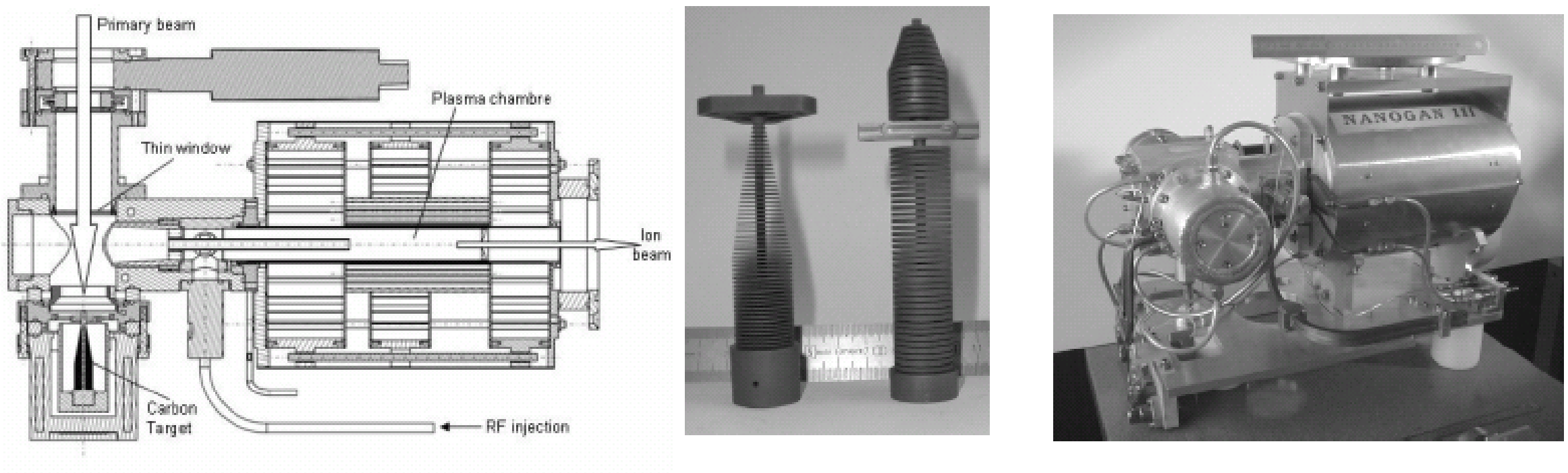

**Fig. 16:** Target and Nanogan III ECR source operated at SPIRAL I, GANIL facility

While these sources have demonstrated the production of noble gases, molecular oxygen, carbon and nitrogen isotope beams, the main factors that affect the ionization efficiencies are the distribution along different charge states, the residence time in a larger plasma volume, and possible losses of the isotopes on the chamber walls. Two different lines of development have since been followed. One is the development of compact 1+, 2.45 GHz ECR sources, such as those reported in Ref. [22]. The second line of development was the implementation of a so-called charge breeding scheme, in which a 1+ beam is injected in an ECRIS or EBIS charge booster or charge breeder device. This second line of development is specifically described in the course given by F. Wenander; the interested reader should therefore refer to his contribution on charge breeding.

### 6.2 Mode of operation and underlying physics principles

These sources are operated following the same principles as their counterpart operated for stable beams. Here also the interested reader is advised to look in publications describing ECR ion sources, one of those is R. Geller Electron Cyclotron Resonance ion Source and ECR Plasmas, (IOP Bristol and Philadelphia, 1996). The main point of interest when applied to the production of radioactive ion beams is the residence time of the isotopes, coupled to the ionization efficiency. Studies have been performed in an attempt to link plasma chamber volume, residence time and ionization efficiency dependent on isotope half-life, as shown in Fig. 17 for Ne isotopes. Obviously more systematic studies are required, especially for molecular beams of short-lived isotopes.

### 6.3 Specific applications, future developments

Development of RF-heated plasma sources for radioactive ion beams is a very active field in our community. Production of radiation-hard ECR sources is progressing well. Furthermore, new concepts of plasma sources have appeared during the past few years, for instance, low-frequency inductively coupled RF sources or Helicon-type RF sources. These compact and robust sources are expected to provide interesting results for the production of different molecular beams, in an attempt to produce plasmas that are more appropriate for the ionization of the molecules without destruction of the molecular bonds [24].

## 7 Conclusion

I hope I have been able to share some of the exciting recent developments that our field has witnessed over the past few years. Many new facilities are under conception or construction around the world. At the same time, new types of radioactive beams are produced each year by the ISOL method, reflecting the progress in the various fields linked to beam production. The family of robust, compact, simple, rapid and efficient sources suited for these facilities will almost certainly continue its expansion.

Progress in the understanding of the basic physical phenomena and the development of new concepts will probably lead to increases in quality and intensity of the beams.

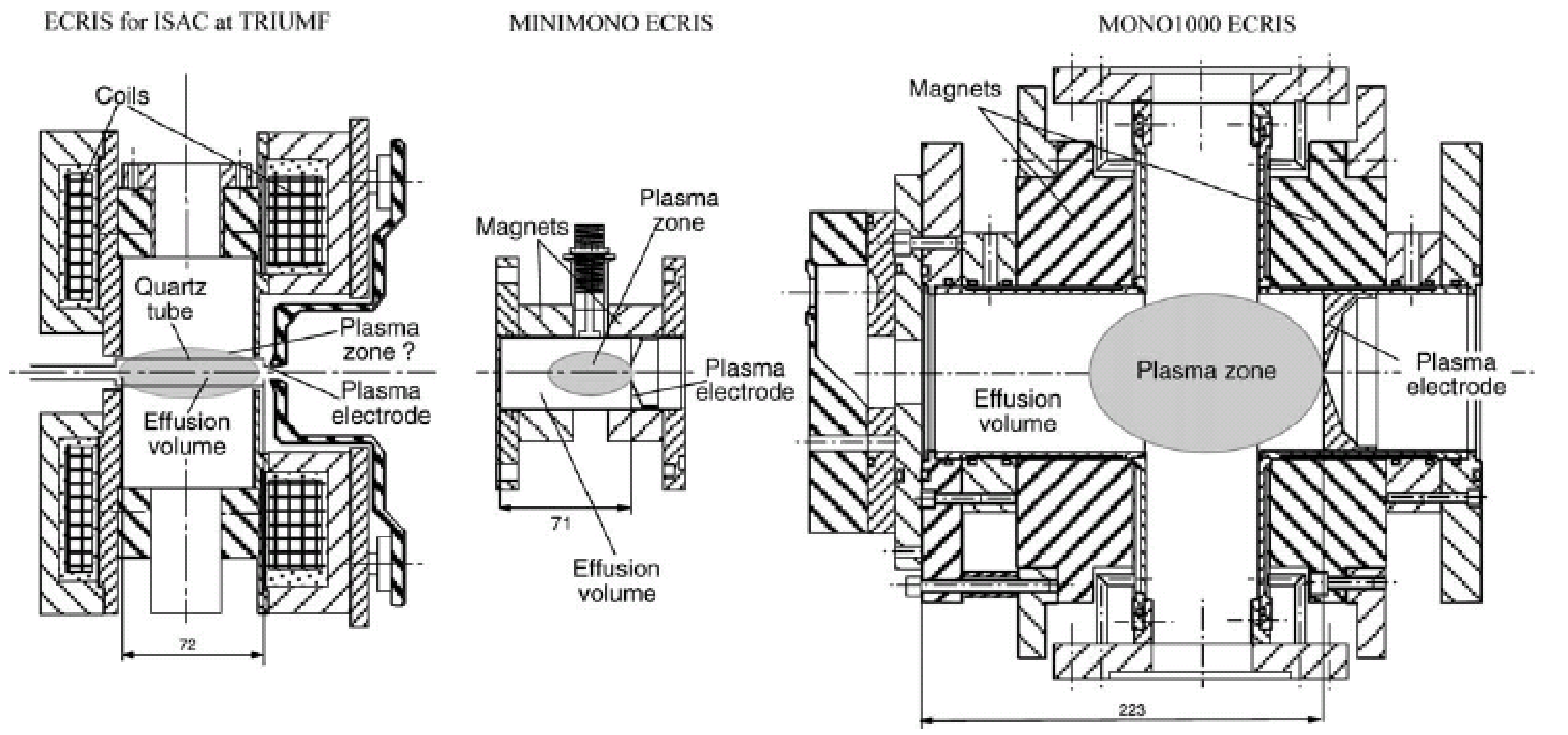

Comparisons between expected and measured ionisation efficiencies with four 2.45-GHz ECR ion sources, for different isotopes of Ne

| | ISAC TRIUMF | MINIMONO GANIL | MONO1000 GANIL | MONOBOB GANIL |
|---|---|---|---|---|
| Volume (cm$^3$) | 18 | **117** | **1500** | **687** |
| Exit diameter (mm) | **5** | **4** | 7 | **5** |
| Transient time (ms) | <**40** | 34 | 15 | 81 |
| Ion. efficiency (%) | | | | |
| $^{20}Ne^+$ (stable) | 11.5 | **25** | **92** | **53** |
| $^{22}Ne^+$ (9.25 s) | **12** | 25 | 92 | 53 |
| $^{18}Ne^+$ (1.67 s) | 11 | 25 | 91 | 51 |
| $^{17}Ne^+$ (0.109 s) | 10.5 | 19 | 81 | 30 |

**Fig. 17:** Comparison of the different 2.45 GHz ECR sources at GANIL and TRIUMF (from Ref. [23])